\documentstyle[twocolumn,aps,prl,epsf]{revtex} 
\input{epsf}
\begin{document}
%\tighten
\title{Mean-field Theory of the Localization Transition}
\author{Ferenc P\'{a}zm\'{a}ndi, Gergely Zim\'anyi and Richard Scalettar}
\address{Physics Department, University 
of California, Davis, CA 95616}
\date{\today}
\maketitle

\begin{abstract} 
A mean field theory of the localization transition for bosonic systems is
developed.  Localization is shown to be sensitive to the
distribution of the random site energies.  It occurs in the presence of
a triangular distribution, but not a uniform one.
The inverse participation ratio, the single site Green's function, 
the superfluid order parameter and the
corresponding susceptibility are calculated, and the appropriate
exponents determined. All of these quantities indicate the presence
of a new phase, which can be identified as the {\it Bose-glass}.


\end{abstract}

\narrowtext

\vskip 0.7cm

The localization transition in disordered systems has been a focus
of statistical and condensed matter physics ever since the famous 
paper of Anderson \cite{phil}. Though
most of the work deals with the metal-insulator transition in
fermionic systems, there is a recent surge of interest in bosonic
models\cite{fisher,rokhsar}. The superfluid-insulator transition in granular
superconductors \cite{hTc} or $^4$He in disordered media \cite{reppy}
are the paradigmatic experimental realizations of such systems. 
The theoretical understanding of phase transitions is typically
based on a mean field description and subsequent fluctuation analysis. 
The generally held belief even today is that
the spatial homogeneity of this ``mean field" allows only for extended
states, thus obliterating the localized phase \cite{fisher}.
Below we demonstrate that in a model of hard core bosons with random site
energies and infinite range hopping this
conventional wisdom does not hold, and with a suitable 
choice of the disorder distribution the localization transition 
can indeed be captured within the mean field technique.

To highlight the involved physics, consider 
the standard argument against a transition in infinite connectivity lattices.
It is believed\cite{fisher} that because of the infinite number of neighbors,
every site will be connected to ``virtually degenerate" sites for 
a continuous distribution of the disordered site energies. 
Thus hopping between these sites {\it always} gains kinetic energy
with zero cost in potential energy, delocalizing the particles.  
However let us observe that for a finite system of size $N$, the
gain from hopping between two sites is ${\cal O}(1/N)$ and the potential
energy difference to the energetically closest sites is of the
same order, since one has $N$ site energies chosen independently from
a finite interval. This means that there is a finite probability
that {\it no sites at all} are available within the $1/N$ energy window 
set by the kinetic energy.  When hopping from such a site,
the potential energy cost certainly outweighs the kinetic energy gain.
A complex sum of gains and costs will decide
whether a state will be localized or extended. 

To demonstrate the nontriviality of this sum, consider the spectrum of the
ordered array. The ground state is homogeneous, non-degenerate
with an energy of -1, clearly maximally benefitting from the kinetic term. 
{\it All} of the other N-1  single particle excited
states possess zero energy. This is so because in this model
to ensure orthogonality to the ground state the wavefunctions
of the excited states have fluctuating signs across the sample. 
This {\it destructive quantum interference} frustrates the kinetic term,
preventing any gain from the hopping process.
So for weak disorder there is more kinetic energy to be gained by
staying extended, and only a sufficiently strong disorder can localize 
the ground state.
For the excited states however staying localized at a suitable site 
offers the lowering of the total energy in the absence
of a kinetic energy premium, thus the excited states will become
localized for {\it arbitrarily small disorder}.
As we will see, this physics can be brought out by a non-traditional
choice of the disorder: we will focus on the triangular distribution of 
the site energies.

To establish the framework of the physics, consider the Hamiltonian
\begin{equation}
H=-\sum_{i,j} J_{ij}a_i^\dagger a_j
%- \sum_i (\mu+h_i)a_i^+a_i,
\label{ham}
\end{equation}
where $a_i^{\dagger}$ ($a_i$) creates (annihilates) a hard-core boson
at site $i$ ($i=1,\dots,N$). We consider the case of infinite range
hopping: $J_{ij}=N^{-1}$ for all pairs, where the mean-field approach is exact.
$J_{ii}=\mu + h_{i}$, where $\mu$ is the chemical potential
and $h_i$ is a random on-site energy.
We recall that \cite{fisher} in the absence of disorder there are two phases
of the model: for generic fillings the bosons can propagate.
At zero temperature they form a superfluid. On the other hand for
precisely one boson per site, the particles localize in
a Mott insulating phase.  When we take away a single particle from 
this insulator, the resulting hole will behave exactly as the first particle
added to the empty lattice. This can be proven by performing a particle-hole
transformation on the Hamiltonian.

In what follows we approach the localized phase from two directions.
First, from the Mott insulator by adding holes. 
In this case a one particle approach clearly suffices.
We calculate the density of states (DOS) and the participation ratio, 
and show that above a critical disorder the ground state becomes localized.
Critical exponents will be evaluated as well. Second, we approach the same
phase by decreasing the density of holes from the extended phase and compute 
the superfluid order parameter and the susceptibility. 
%and particle density.

In studying the one particle problem, the quenched disorder averaging
$<\cdots>_{ave}$ is performed with the replica trick, allowing us to rewrite 
the DOS as:
\begin{eqnarray}
\rho(\lambda)&=& {1\over N} <\sum_{k}\delta(\lambda+J_{k})>_{ave}\\
\nonumber
&=&{\rm lim}_{n\rightarrow 0}{2\over N\pi} {\rm Im} 
{\partial\over\partial \lambda}
{\partial\over\partial n}\int [dx_{i\alpha}] ~d{\bf h} P({\bf h})\\
\nonumber
&{\rm exp}&{\bigl(}-{\lambda\over 2} \sum_{i\alpha}x_{i\alpha}^2-{1\over 2} \sum_{ij\alpha} J_{ij} x_{i\alpha}x_{j\alpha} {\bigr)} ~~.
\label{rho}
\end{eqnarray}
 $\lambda$ is an energy eigenvalue with a positive infinitesimal imaginary part
and the $J_{k}$'s are the
eigenvalues of the $J$ matrix. Next the off-diagonal terms
are decoupled using an auxiliary field $z_{\alpha}$,
and the $x$ integrals are performed to transform the exponent into:
\begin{eqnarray}
S_{eff}&=&-\frac{1}{2}\sum_{\alpha}~z_{\alpha}^2 + N ~ \ln
<\exp ~{\bigl [}-\frac{n}{2}~\ln~(\lambda - h){\bigr ]}\\
\nonumber
&-& \frac{1}{2N} 
\frac{1}{(\lambda - h)}\sum_{\alpha}~z_{\alpha}^2{\bigr ]}> ~~ ,
\end{eqnarray}
where the averaging over the disordered 
site energies $\int dh P(h) A(h)$ is denoted by $<A(h)>$.
In the expectation value we use the inverse of the number of lattice
sites $N$ as a small parameter to carry out an exact expansion. We also
keep only the terms linear in the replica number $n$ to arrive at the exponent:
\begin{eqnarray}
S_{eff}=-\frac{1}{2}{\Bigl (}[1+<\frac{1}{\lambda -h}>]
\sum_{\alpha}z_{\alpha}^2 + Nn<\ln (\lambda - h)>{\Bigr )}
\nonumber
\end{eqnarray}
Finally performing the integration over the auxiliary field $z_{\alpha}$ yields:
\begin{eqnarray}
\rho (\lambda)&=& -\frac{1}{\pi} {\rm Im} <\frac{1}{\lambda - h}>-\frac{1}{N\pi}
{\rm Im} \frac{\partial}{\partial \lambda}\ln (1+<\frac{1}{\lambda - h})>
\nonumber\\
&=& P(\lambda) +\frac{1}{N} \delta(\lambda+\lambda_0),
\label{ro}
\end{eqnarray}
where $\lambda_0$ is obtained from: $<(\lambda_0-h)^{-1}>=1$.  
We assume that $P(h)$, the distribution of the site energies is non-zero on 
the interval $(-\Delta,\Delta)$. 

The DOS was easiest to obtain in the above path-integral framework, but
the subsequent physical quantites can be determined by the simpler method of
calculating the eigenvectors ${\bf \varphi}(\lambda)$ of the $J$ matrix.   
One finds $\varphi_{i}(\lambda)=m/(\lambda-h_{i})$, where $m$ is the superfluid
order parameter: $m=(1/N) \sum \varphi_{i}$. 
The self-consistency equation for $m$ yields:
$1=1/N \sum 1/(\lambda-h_{i})$.  If $\lambda$ is inside $P(h)$, then
a few $\varphi_{i}(\lambda)$'s will be $\sim {\cal O} (1)$ and the rest of the
$\varphi_{i}(\lambda)$'s will be $\sim {\cal O} (1/N)$ to satisfy the 
normalization condition, i.e. the states inside the continuum are localized,
whereas for $\lambda_{0}$ outside the continuum {\it all} 
$\varphi_{i}(\lambda_{0}) \sim 
{\cal O} (1/{\sqrt N})$, describing an extended state.  
This picture is in complete accordance with the above determined DOS.
The ground state becomes localized if $-\lambda_{0}$ reaches the bottom 
of the band. This does not happen for a rectangular distribution 
since explicit calculation yields
$\lambda_0=\Delta \coth \Delta >\Delta$.
However for the triangular distribution 
$P(h)=\Delta^{-2}(\Delta-|h|)$ the ground
state eigenvalue $-\lambda_0$ 
is given implicitly by the equation
\begin{equation}
\Delta^2=
(\lambda_0-\Delta)\ln\frac{\lambda_0-\Delta} {\lambda_0}+
 (\lambda_0+\Delta)\ln\frac{\lambda_0+\Delta} {\lambda_0},
\label{triangle}
\end{equation}
which is valid only for $\lambda_0>\Delta$.  Taking the
$\lambda_0\rightarrow \Delta+0$ limit in Eq. (\ref{triangle}) shows that
$-\lambda_{0}$ reaches the bottom of the band at the critical disorder 
$\Delta_c=2\ln 2 =1.38$. For $\Delta > \Delta_{c}$
the DOS becomes identical 
to the distribution of the site energies, i.e. $\rho({\lambda})=P(\lambda)$.

One measure of localization is the
participation ratio ${\cal P}=(\sum_i |\varphi_i|^2)^2/(\sum_i |\varphi_i|^4)$.
${\cal P}$ is proportional to the system size $N$ for extended states
and remains $\sim {\cal O} (1)$ for localized ones.  
The expression for $\varphi_{i}$ yields
\begin{equation}
{{\cal P} \over N}=\frac{ <(\lambda_0-h)^{-2}>^2} 
                 { <(\lambda_0-h)^{-4}> },
\label{p/N}
\end{equation}
for $\Delta<\Delta_c$ and ${\cal P}/N=0$ for $\Delta>\Delta_c$. 
For the uniform distribution 
there is no critical disorder, and the participation ratio is
${\cal P}/N=3/(3\cosh^2\Delta+\sinh^2\Delta) \sim {\cal O} (1)$,
clearly indicating that the ground state remains extended for any finite
disorder strength. 
For the triangular distribution the
result of the integral (\ref{p/N}) is
\begin{equation}
{{\cal P} \over N}={{3 \lambda_0^2}\over {3\lambda_0^2-\Delta^2}}
\biggl({{\lambda_0^2-\Delta^2}\over {\Delta^2}}\biggr)^2
\ln^2{{\lambda_0^2-\Delta^2}\over{\lambda_0^2}},
\label{ptri}
\end{equation}
and $\lambda_0$ is obtained from Eq. (\ref{triangle}) for $\Delta<\Delta_c$. 
As $\lambda_0$ approaches $\Delta$,
${\cal P}/N$ disappears as $\propto(\Delta_c-\Delta)^2$, indicating that for
$\Delta>\Delta_c$ the ground state becomes {\it localized}. From now on 
we will focus only on this more interesting case of
the triangular distribution.

\begin{figure}[t]
\epsfxsize=3.3in
\epsfysize=2.5in
\epsffile{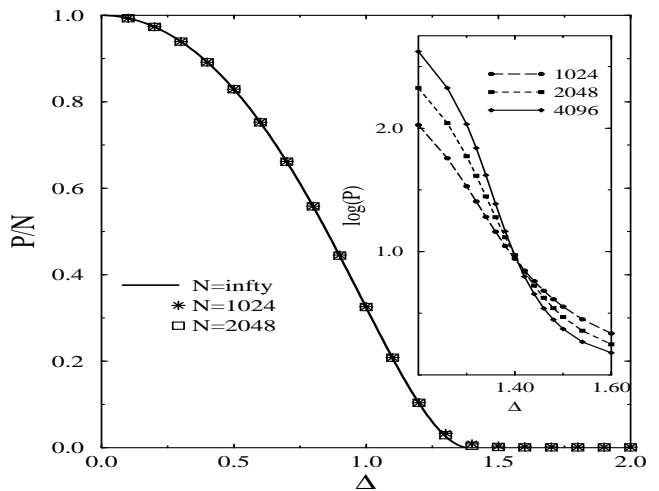}
\vskip 0.2cm
\caption{Analytical and numerical results for the ground state participation
ratio.}
\end{figure}

One can independently test these results numerically by
diagonalizing the matrix $J$ and measuring $\cal P$.
Fig.~1 displays convincing agreement between analytic results (solid line)
and numerical results (symbols).  
The crossing and subsequent decrease of ${\cal P}$ with $N$ (inset) provides
compelling evidence for the localization of the ground state.
To demonstrate how ${\cal P}$ distingishes between extended and localized 
states, one usually considers two possibilities, $\varphi_{i} = 1/\sqrt{N}$
and $\varphi_{i} = \delta_{i,i_{0}}$.  These yield
${\cal P}=N$ and ${\cal P}=1$ respectively.
One often equivalently phrases this analysis in
terms of the inverse participation ratio ${\cal P}^{-1}$ which
takes on small values ${\cal O}(1/N)$ for extended states and values near unity
for localized ones.
However, if $\varphi_{i}$ has a few
large amplitudes on selected sites, and an extended background,
$\varphi_{i} = [1/2] \delta_{i,i_{0}} + 1/\sqrt{2(N-1)}(1-\delta_{i,i_{o}})$,
${\cal P}$ and ${\cal P}^{-1}$ will be close to one
and indicate, incorrectly, that the state is localized.
To eliminate the possibility of misidentifying
such ``pseudo-localized'' states, we computed 
${\cal P}^{-1}(n)$, which we define by
systematically removing the $n$ sites where the wavefunction
assumes its largest values.  
%Our mean field lattice has a trivial spatial topology, 
%and this is the simplest analog of the usual technique of 
%monitoring localization by looking at the spatial
%decay of the wavefunction.
We show the results of this calculation in Fig.~2.

\begin{figure}[t]
\epsfxsize=3.2in
\epsfysize=2.5in
\epsffile{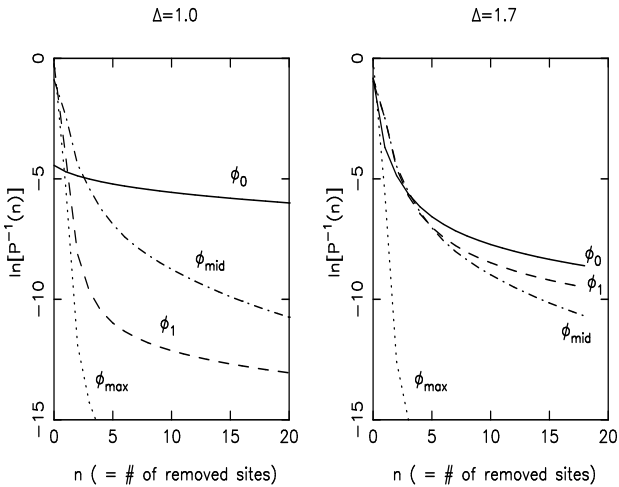}
\vskip 0.3cm
\caption{The truncated participation ratio. Here the four indices refer
to the ground state, first excited state, a mid-band and the maximal energy
wavefunctions.}
\end{figure}

When $\Delta=1.0 < \Delta_{c}$, the ground state is extended, and 
${\cal P}^{-1}(n)$ remains ${\cal O}(1/N)$ as sites are removed.  
Meanwhile, for all the excited states,
${\cal P}^{-1}$ rapidly plunges to much smaller values than
${\cal O}(1/N)$, emphasizing that as a few sites are removed from the
sum, the weight of the remaining sites is negligible.
%Had the states been ``pseudo-localized",
%{\cal P}$^{-1}$ would have fallen rapidly from 1, but then
%saturated at ${\cal O}(1/N)$.
We checked this behavior on different lattice sizes,
and found that ${\cal P}^{-1}(n)$ fell slightly {\it more}
rapidly with $n$ for larger lattices, eliminating the possibility
that the states were extended but over a small fraction of the lattice.
We conclude that all states in the continuum are indeed localized.
When $\Delta=1.7 > \Delta_{c}$ even the ground
state is localized, as shown.

Returning to the analytic calculations,the transition can be parametrized 
by the chemical potential as well.
If $\mu$ exceeds a critical value $\mu_c$, then the ground state 
has no holes. At $\mu_c$ the first hole appears in the system.
$\mu_{c}$ is given by the lowest eigenvalue of the $-J$ matrix, i.e.
$\mu_c=\lambda_0$ for $\Delta<\Delta_c$ and $\mu_c=\Delta$ for
$\Delta>\Delta_c$.
The transition is well captured by the imaginary time on-site Green's function
\cite{PD}
\begin{equation}
g(\tau)\equiv\frac{1}{N}\sum_i \langle T_{\tau} 
a_i^{\dagger}(\tau) a_i(0)\rangle _{0}=
\int d\lambda \rho(\lambda) \exp[-\tau(\mu-\lambda)],
\label{gtau}
\end{equation}
for $\tau>0$, where $\langle\cdots\rangle _{0}$ is the ground state 
expectation value, and $T_{\tau}$ is the time ordering operator. 
For weak disorder the transition happens to the state at 
$-\lambda_0$. Combining Eqs. (\ref{ro}) and (\ref{gtau})
one obtains {\it at criticality}: 
$g(\tau)\approx 1/N + \exp[-\tau(\lambda_{0}-\Delta)]$
near the transition, for $\tau\gg 1$. On the other hand, 
for strong disorder ($\Delta>\Delta_c$) the critical $g(\tau)$ decays as:
$g(\tau)\propto \tau^{-2}$.
%Though in both cases the gap $\mu-\mu_c$ disappears linearly,
%the imaginary time correlations right at the critical point are 
%different. 
This change of the critical behavior suggests that we 
enter into different phases for  $\Delta<\Delta_c$ and $\Delta>\Delta_c$.
It is worth noting that if the distribution
$P(h)\propto (\Delta-h)^{\alpha}$, then in the strong disorder regime
$g(\tau)\propto \tau^{-\alpha-1}$ right at the critical point.
This means that in the strong disorder regime the critical
exponent depends on the distribution of the disorder.

If one approaches the same localized phase from
higher densities of holes, these one particle techniques must be abandoned.
We will concentrate on the free energy of the Hamiltonian (\ref{ham}):
$f=k_{B}T {\rm ln}Z /N$, where $Z={\rm Tr} \exp -\beta H$.
Introducing the magnetization $m$ as a Hubbard-Stratonovich field
decouples the different sites to yield a single site problem:
\begin{equation}
f=\frac{1}{4} m^2 - {1\over \beta N} \sum_{i} {\rm ln} Q_{i},
\end{equation}
where
\begin{equation}
Q_{i}= {\rm Tr} ~\exp ~\beta [(\mu+h_{i}) ~a^{\dagger} a + 
m a^{\dagger} + m^* a]
\end{equation}
For hard-core bosons the Hilbert space is only two dimensional, as a site may be
only empty or occupied by one particle. This allows the exact
evaluation of $Q$ and hence the free energy per site, which we give here 
only for $T=0$:
\begin{equation}
f=\frac{1}{4} m^2 - \frac{1}{2}<(\mu-h)+\sqrt{(\mu-h)^2+m^2}>.
\end{equation}
$m$ is determined by the saddle point condition:
$m=m <[(\mu-h)^2+m^2]^{-1/2}> $.
The order parameter $m$ is proportional to $\langle a_i\rangle$. 
The hard-core boson problem is
equivalent to a spin 1/2 XY model in a transverse field, and $m$ 
corresponds to the magnetization in the XY plane.

$m=0$ is always a solution, but upon exiting the Mott insulator
below a critical chemical potential $\mu_c$ another
solution with nonzero magnetization $m$ appears. 
For weak disorder $\Delta<\Delta_c$ $\mu_c=\lambda_0$. 
Expanding in $m$ near the critical
point the magnetization behaves the same way as it does without disorder, i.e.
$m\propto \sqrt{\mu_c-\mu}$, the well-known Landau result.

In the strong-disorder regime when increasing the density, the particles 
occupy localized states. Do we have a superfluid in
this case? In order to get the answer, we write out the
saddle-point equation for the triangular distribution:
\begin{eqnarray}
\Delta^2&=&\mu_+
\ln\biggl({\mu_++\sqrt{\mu_+^2+m^2}\over \mu+\sqrt{\mu^2+m^2}}\biggr)\\
\nonumber
&+&\mu_-
\ln\biggl({\mu_-+\sqrt{\mu_-^2+m^2}\over \mu+\sqrt{\mu^2+m^2}}\biggr)\\
\nonumber
&+& 2\sqrt{\mu^2+m^2}-\sqrt{\mu_+^2+m^2}-\sqrt{\mu_-^2+m^2},
\label{mtri}
\end{eqnarray}
where we have introduced the notation $\mu_+=\mu+\Delta$ and
$\mu_-=\mu-\Delta$. For strong disorder $\mu_c=\Delta$, because if
$\mu$ is slightly below $\Delta$, i.e. $\mu_-$ is negative, one sees 
immediately from the second term of the right hand side (RHS)
of Eq. (13) that $m$ must be different from zero,
otherwise the logarithm 'blows up'. Clearly a superfluid is formed,
even though the constituent particles possess localized wavefunctions.
The physics is then that the {\it phases} of the localized wavefunctions
lock up to yield the off diagonal long range order.
This situation is very reminiscent to the ``localized superconductor"
picture of Ma and Lee \cite{malee}. 

The above results clearly show that the truly localized phase occupies 
only a line in the $\mu - \Delta$ plane. However we think that this is indeed
the seed of the ``Bose-glass" phase, because for the finite
range hopping model if one partitions the system into blocks of the
size of the hopping length, each will support only one of these truly
localized, non-superfluid states.
There will be a macroscopic number of these blocks, thus the original
line will expand into a finite region as a function of the density. 

The one-particle states being localized, it is
natural to assume that $m\ll (\mu_c-\mu)=|\mu_-|$. In this
case the second term of the RHS of Eq. (13) behaves
like $\mu_-\ln(m^2/|\mu_-|)$, and this term must be finite,
so we find
\begin{equation}
m\propto \sqrt{\mu_c-\mu}\exp\Bigl(-{a\over \mu_c-\mu}\Bigr),
\label{m}
\end{equation}
where $a=\Delta(\Delta-\Delta_c)/2$. As we can see, below $\mu_c$
the system is superfluid, though the magnetization is much smaller
than in the weak-disorder case. The critical behaviour is again
different in the weak and strong disorder regime.
% as in the latter case the magnetization exhibits an essential singularity.
It is remarkable that in the study of the corresponding one dimensional
problem also essential singularities were found \cite{giam}.

%Explicit knowledge of the free energy allows us to calculate
%other observables as well. The density of particles is given by
%$n=-\frac{\partial f} {\partial \mu}$. Direct evaluation yields
%$n \propto (\mu_{c}-\mu)^2$ for low densities. This form indicates that
%every state is occupied by a single hard core boson, 
%in accordance with the fact that the eigenstates are localized.

Finally we calculate the susceptibility by adding an infinitesimal in-plane
field $B$ to $m$ and taking $\chi=- (\partial ^2 f / \partial B ^2)$.
Approaching the transition from the $m>0$ side
for arbitrary disorder we get $\chi \propto (\mu-\mu_{c})^{-1}$.
On the other hand approaching from the non-superfluid side (i.e. $m=0$),
while for weak disorder one again obtains $\chi \propto (\mu-\mu_{c})^{-1}$,
for strong disorder $\chi$ remains 
finite even {\it at} the transition. This difference of the exponents again
demonstrates that the transitions from the Mott insulator into 
the superfluid or into the localized region belong to different universality
classes, further strengthening the argument that the localized region
in fact is a separate phase. Just as for the Green's function, the above
exponents also depend on the asymptotics of the disorder distribution
for strong disorder.

To summarize, we investigated the disordered hard core boson problem.
We proved that, contrary to previous beliefs, the mean field theory is able 
to capture the localization transition. We approached the localized region
both from the Mott insulator and from the superfluid phase, calculating
the density of states, the inverse participation ratio, the on-site Green's
function, the magnetization and the susceptibilty. We tested our theory with
independent numerical investigations and found detailed agreement.
The important observation is that the exponents of all of the above physical
quantities {\it differ} for the direct insulator - superfluid and the 
insulator - glass or superfluid - glass transitions, clearly demonstrating
that the glass transition belongs to a new universality class.
We also gave a real space blocking argument why we expect the glass to expand
into a region in the parameter space when the range of hopping
is reduced to finite values.

We would like to acknowledge useful discussions with Tom Devereaux
and Kyungsun Moon.
This work has been supported by NSF-DMR-92-06023, and by the US-Hungarian 
Joint Fund 265/92b.


\begin{references}
\bibitem{phil}
P. W. Anderson, Phys. Rev. {\bf 109} 1492 (1958).

\bibitem{fisher}
M. P. A. Fisher, P. B. Weichman, G. Grinstein, and D. S. Fisher, 
Phys. Rev. B {\bf 40}, 546 (1989).

\bibitem{rokhsar} M. Wallin, E.S. Sorensen, S.M. Girvin and
A.P. Young, Phys. Rev. {\bf B49}, 12115 (1994);
K. Singh and D.S. Rokhsar, Phys. Rev. {\bf B46}, 3002 (1992);
P. Nisamaneephong, L. Zhang and M. Ma, Phys. Rev. Lett. {\bf 71}, 3830 (1993).

\bibitem{hTc}
A. F. Hebard and M. A. Palaanen, Phys. Rev. Lett. {\bf 65}, 927 (1990); 
H. M. Jaeger, D. B. Haviland, A. M. Goldman, 
and B. G. Orr, Phys. Rev. {\bf B34}, 4920 (1986).

\bibitem{reppy}
D. Finotello, K. A. Gillis, A. Wong, and M. H. W. Chan, 
Phys. Rev. Lett. {\bf 61}, 1954 (1988); 
J. D. Reppy, J. Low Temp. Phys. {\bf 87}, 205 (1992). 

\bibitem{PD}
F. P\'azm\'andi and Z. Doma\'nski, to appear in Phys. Rev. Lett.

\bibitem{malee}
M. Ma and P. A. Lee, Phys. Rev. {\bf B32}, 5658 (1985).

\bibitem{giam}
T. Giamarchi and H. J. Schulz, Phys. Rev. {\bf B37}, 325 (1988).

\end{references}
\end{document}